# Constructing AI ethics narratives based on real-world data: Human-AI collaboration in data-driven visual storytelling

Mengyi Wei
mengyi.wei@tum.de
Technical University of Munich
Munich, Bavaria, Germany

Chenjing Jiao
cjiao@ethz.ch
ETH Zürich
Zürich, Switzerland

Chenyu Zuo*
chenyu.zuo@csfm.ethz.ch
ETH Zürich
Zürich, Switzerland

Lorenz Hurni
lhurni@ethz.ch
ETH Zürich
Zürich, Switzerland

Liqiu Meng
liqiu.meng@tum.de
Technical University of Munich.
Munich, Bavaria, Germany

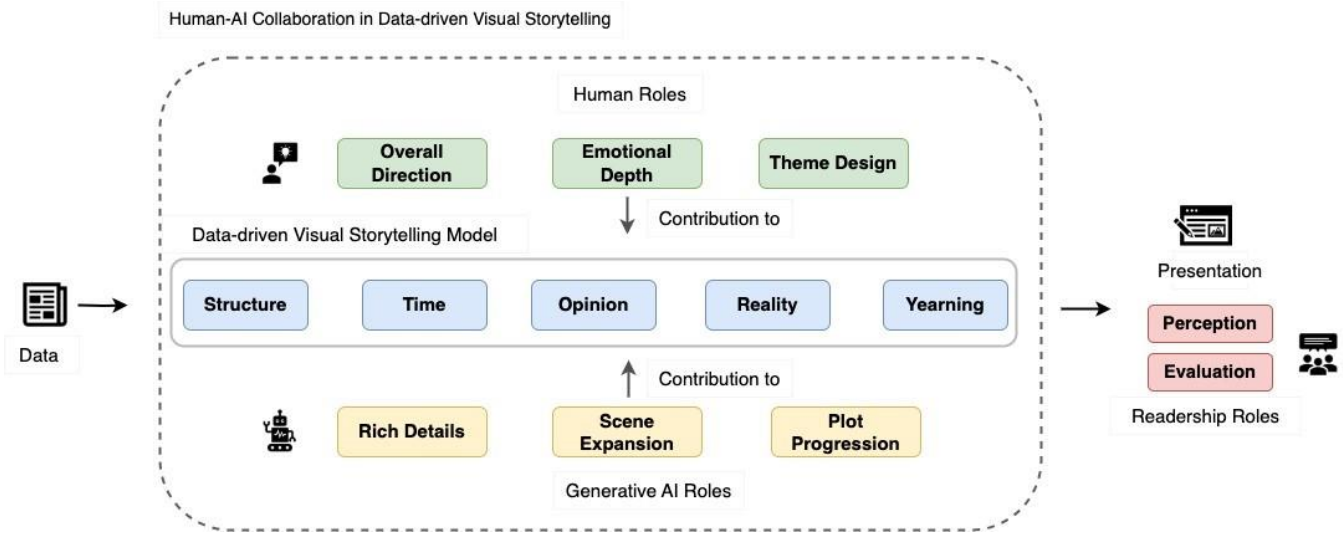

Figure 1: A framework for human-AI collaboration in data-driven visual storytelling.

## Abstract

AI ethics narratives have the potential to shape the public's accurate understanding of AI technologies and promote communication among different stakeholders. However, AI ethics narratives are largely lacking. Existing limited narratives tend to center on works of science fiction or corporate marketing campaigns of large technology companies. Misuse of "socio-technical imaginary" can blur the line between speculation and reality for the public, undermining the responsibility and regulation of technology development. Therefore, constructing authentic AI ethics narratives is an urgent task. The emergence of generative AI offers new possibilities for building narrative systems. This study is dedicated to data-driven visual storytelling about AI ethics relying on the human-AI collaboration. Based on the five key elements of story models, we proposed a conceptual framework for human-AI collaboration, explored the roles of generative AI and humans in the creation of visual stories. We implemented the conceptual framework in a real AI news case. This research leveraged advanced generative AI technologies to provide a reference for constructing genuine AI ethics narratives. Our goal is to promote active public engagement and discussions through authentic AI ethics narratives, thereby contributing to the development of better AI policies.

## 1 Introduction

With the rapid advancement of artificial intelligence (AI) technology, the resulting ethical issues, such as algorithmic bias, privacy violations, accountability, and automation of professional tasks, have affected more and more aspects of society. Raising public awareness is necessary to address AI ethics issues [18]. However, AI ethics often involves complex concepts from technology, law, and philosophy, making it difficult for the public to find appropriate channels to properly understand these issues. This lack of understanding could result in the loss of public voices in technological development, ultimately compromising the regulated and responsible growth of AI technology. At this critical juncture, AI ethics narratives has emerged as one of the remedies.

Human beings, as narrative beings, understand their lives, construct their worlds, and connect through narratives [6, 21]. Narratives are often multidimensional, capable of exploring diverse voices and perspectives, giving voice to individuals and groups, and drawing attention to marginalized or overlooked voices [30]. Most of the existing AI ethics narratives stem from science fiction works or corporate marketing by major technology companies. Hermann cautions against fictional AI narratives, such as those in science fiction, as the public might use them as significant benchmarks for evaluating technology, distorting perceptions of real-world AI and fostering unwarranted fears and concerns [29]. Hudson et al. acknowledge that AI narratives can facilitate the development of better AI policies but argue that relying on science fiction to explore policy-making overlooks pressing and mundane ethical issues in the real world, overly focusing on speculative themes from fictional universes [32]. A growing number of scholars have recognized the dearth of AI ethics narratives and have begun to explore how to create compelling AI ethics narratives [15].

This paper proposes to construct data-driven visual stories based on real-world AI ethics cases. These stories aim to improve public awareness of AI ethics issues and facilitate dialogue among diverse stakeholders to support the development of better policies. Data-driven visual stories combine images and text to convey compelling narratives [48, 49]. Using panel layouts and integrating narrative content, they can communicate insights embedded in data and thus [4, 61], establish an emotional connection between the audience and the data. This method of communicating AI ethics information is clear and accessible to both expert and non-expert readers [5, 60].

However, creating effective data stories requires a combination of skills, including technical expertise in data analysis and transformation, and narrative and artistic design skills. Recent studies have extensively explored the potential of artificial intelligence (AI) to support and enhance human storytelling in data-driven contexts. Most of these studies focus on AI's assistance in specific aspects of data storytelling, such as identifying data patterns [59], organizing numeric-based facts [52], generating infographics [45], or accessing visualization such as in terms of aesthetics [22]. Few studies demonstrate practical workflows that apply real-world data to showcase the collaboration between generative AI and humans in creating data stories. The challenges are manifold, regarding: 1) data sources for constructing authentic AI ethics narratives; 2) models for creating data-driven visual storytelling; 3) human-AI collaboration; and 4) the aesthetic and emotional impact of the story.

Facing these challenges, we developed a framework for human-AI collaboration in data-driven storytelling illustrated in Figure 1. The data used in our research is from an AI Incident Database. We divided the data-driven visual storytelling model into five key elements: structure, time, opinion, reality, and yearning. We assigned different roles played by humans and generative AI within this storytelling model. These roles help to illustrate how humans and AI leverage their strengths to contribute to effective storytelling, demonstrating the feasibility of human-AI collaboration.

Our research aims to create authentic AI ethics narratives based on the human-AI collaboration. The main contributions of this study are as follows: 1) we proposed a human-AI collaboration framework in transforming fragmental texts of certain events into understandable narratives. This lays groundwork for future exploration of human-AI collaboration in understanding and communicating complex information. 2) We showcase AI ethics narratives with real-world news report data, avoiding exaggeration or neglect of the genuine AI ethics issues people face daily. Through AI ethics storytelling, we help the public understand complex AI ethics problems and explore possibilities for improved AI policies.

## 2 Related Work

## 2.1 AI Ethics Narratives

Narrative is one of the oldest forms of human communication, documentation, and entertainment. It involves sequences of events with characters and settings that aid memory, engage audiences, and facilitate understanding [71]. In modern society, the characteristics of narrative are important in helping people navigate an increasingly complex world. In the face of unresolved AI ethics issues, more and more experts and scholars are recognizing the importance of AI ethics narratives. A Royal Society study highlights the urgent need to take AI narratives seriously to improve public reasoning and narrative evidence [17]. They emphasize the importance of 'listening to stories' - actively engaging with and anticipating narratives as a form of public participation, where imaginaries of AI promote public reasoning and inform policy-making [15]. Marmolejo-Ramos et al. propose the construction of AI narratives as a pragmatic approach to promoting public engagement with digital governance initiatives. They aim to increase public awareness and understanding of AI and promote the safe adoption of human-centered and trustworthy AI [43]. Coeckelbergh advocates for narrative responsibility in AI, suggesting that narrative responsibility serves as a means of explaining AI, helping the public understand it, and opposing it when necessary [16]. Watson et al. argue that competing narratives are essential for advancing contemporary AI ethics discourse [62].

However, mainstream AI ethics narratives are dominated by large tech companies, mass media, and science fiction [15]. Mainstream media reinforces public perceptions of AI as "terrifying robots" [9, 10]. Science fiction has become a reference point for scientific communication and public and media discussions about the ethics, biases, and risks of AI [29], serving as a corpus for examining public acceptance of AI and its influence on political decision-making. Movies and science fiction present social relationships in dramatic rather than technical terms, reinforcing views of AI "uprisings" or subtle manipulations by AI agents. Furthermore, using AI ethics narratives derived from science fiction to inform AI policy overlooks ordinary yet real ethical issues, overly fixating on dramatic conflicts [32]. Chubb et al. found that mainstream narratives distract and mislead public understanding and perceptions of AI, highlighting the need for greater attention to missing AI narratives [15]. Our research is based on real-world AI ethics cases to address this gap, aiming to reveal more realistic and accurate events. This approach helps the public build a correct understanding of AI ethics, promotes public awareness, encourages developer reflection, and fosters collective societal responses.

## 2.2 Generative AI for Visualization

Generative AI (GenAI) refers to a subset of artificial intelligence models that can create new and meaningful content, such as text, images, code. The characteristic that sets GenAI apart from other AI models is that it actively produces original outputs based on patterns it has learned from existing data. It can generate new content that often closely resembles real-world data but is unique. In many cases, content created by GenAI can hardly be distinguished from human craftsmanship [20].

Large language models (LLMs) are advanced GenAI models, which are originally designed to understand and generate humanlike text. LLMs are "large" because of enormous numbers of parameters (or weights) they have, which are adjusted during the training process to allow the models to learn patterns in the data. They are trained on diverse and extensive datasets containing text from books, news, social media, etc., enabling them to generate coherent and contextually relevant responses for various types of tasks and topics, such as answering questions, translating languages, summarizing documents, and even creating conversational responses [11][64].

LLMs are usually based on Transformer architectures, which is a type of neural network especially suited for handling sequential data (like language). The Transformer architecture relies on attention mechanisms, which allocate varying weights to input elements, allowing the model to focus on the most relevant information [56]. Transformer is proficient in establishing global semantic information interaction and processing all parts of a text sequence simultaneously, which enables it to generate coherent and context-sensitive text [67].

Transformer is the key innovation behind the Generative Pretrained Transformer (GPT) developed by the AI research organization OpenAI [1]. GPT is designed to generate human-like coherent and context-aware text in response to prompts [40]. It is pre-trained on vast datasets of text before being fine-tuned for specific downstream tasks. GPT series include GPT-3, GPT-4 and GPT-4o mini[2]. GPT-3 has over 175 billion parameters, while GPT-4 has even more [1]. GPT has been applied and reshaped numerous fields including content generation, automated customer support and code completion and debugging. Notably, GPT has been embedded in research workflows for tasks such as automated literature review, data summarization and data visualization [3][28][34].

Recently, GPT-4 combines LLMs with other modalities. Thus, it can handle both text and images, and is able to process text prompts and provide image-based answers, blurring the line between traditional LLMs and other types of GenAI [1]. Developers have developed customized versions of GPTs specialized in specific tasks (e.g., visualization, coding support) [23]. Based on GPT-4, the Image Generator[3] (IG) is developed by NAIF J. Alotaibi specially for image synthesis from text prompts.

In a broad sense, visualization is the process of creating images, diagrams, or animations to communicate and present information in a way that facilitates human cognition. In the context of data science, visualization involves creating charts, graphs, maps, and other images to reveal patterns, trends, or insights from raw data, which assists exploratory data analysis [69]. A lot of academics and practitioners have tried to leverage and integrate GenAI's generative capacity into visualization frameworks [33].

Natural Language to Visualization (NL2VIS) accepts and interprets user's natural language queries. Then, based on the queries it produces visualizations that align with the user's intent [51][69]. Wu et al. (2024) demonstrate the effectiveness of utilizing LLMs for automating data visualization from natural language descriptions and the performance of GPT-4 stands out. Maddigan and Susnjak (2023) verify that LLMs together with appropriate prompts offer a reliable approach for rendering visualizations from natural language queries. Specially, ChatGPT is demonstrated to have superiority in transforming natural language inputs into more informative and readable visualizations than other models. Although relying on a prompt-based method may not effectively enable LLMs to fully grasp the complexities of the NL2VIS task, prompt-finetuning in a conversational way helps LLMs to iteratively generate an optimal output [66][42].

Narrative visualization presents data or information in a way that integrates storytelling elements to help audiences understand and engage with complex datasets. It combines the analytical power of data visualization with the compelling nature of narratives to create meaningful and memorable communication. Usually, narrative visualization has a clear storyline or sequence [37][13]. Ferracani et al. (2024) present and develop a mobile application that achieves interactive narrative storytelling based on GPT-4 with AI generated illustrations. This application helps improve cultural tourism experiences by leading users in the cities of Rome, Florence and Venice [19].

## 2.3 Human-AI Collaboration in Data-driven Visual Storytelling

Storytelling uses voice, sound, and visuals to convey events, often involving characters and settings [26]. Data-driven visual storytelling is a process that centers on data, utilizing visual techniques such as charts, maps, and animations to reveal insights behind the data while guiding the audience to understand, explore, and act through a narrative approach. It integrates data analysis, visual design, and storytelling techniques to transform complex

---

data into easy-to-understand and engaging visual representations [71]. Segel and Heer first introduced the concept of storytelling with data, summarizing the design space of narrative visualization through a review of existing examples [49]. Creating an effective "data story" requires a diverse skill set. Compelling storytelling requires familiar skills for filmmakers rather than relying solely on technical expertise in computer engineering and science [24]. Lan et al. explored how data stories evoke emotional engagement by introducing Kineticharts to accurately convey emotions, enhancing the expressiveness of data stories and increasing user participation [36]. Li et al. focused on how to discover insights (explore data), transform those insights into narratives (create stories), and deliver the narratives to audiences (tell stories) [37]. These studies often focus on specific aspects of data-driven visual storytelling, such as analyzing data or enhancing emotional appeal, but fail to address how to effectively craft data stories across a multistage pipeline.

More researchers have introduced artificial intelligence to help storytelling to meet the demand for crafting compelling data-driven visual stories. For example, AI can help humans identify interesting data facts, organize story fragments, generate visualizations, or provide feedback [38]. However, these studies fail to comprehensively understand how humans and AI collaborate in data-driven visual storytelling. Chen et al. examined how automation is gradually integrated into visualization design and narrative processes through narrative visualization tools, enabling users to create narrative visualizations more easily [13]. Li et al. conducted a literature review and classified fine-grained collaboration models in storytelling tools, identifying common collaboration patterns in existing tools [38]. While these studies describe human-AI collaboration in storytelling from an overall perspective, they primarily focus on tools. Although tools are essential for creating visual stories based on data, the key lies in the elements that make up a compelling story. To address this gap, our research begins with the model of data-driven visual storytelling and introduces five critical elements of storytelling that balance data science and visual design. More specifically, we explore the roles of humans and AI within the elements of data-driven visual stories. Our framework is designed to be applied to real-world AI news cases, providing a fine-grained demonstration of how humans and AI collaborate to build effective visual stories and providing valuable insights for achieving AI ethics narratives.

## 3 Methodology

This section first introduces a framework for representing human-AI collaboration in data-driven visual storytelling and demonstrates the roles humans and AI play in each storytelling element. Then it applies this framework to a real news case to achieve human-AI collaboration in the construction of AI ethics narratives.

### 3.1 Framework

Telling an effective data-driven visual story requires various skills, including narrative techniques and artistic design [49]. Based on a literature review, a good story should meet the following five criteria: structural clarity, temporal constraints, insightful perspective, realistic plot, and aspirational emotion [12, 27, 44, 63].

This paper summarizes the model of data-driven visual storytelling into five key elements: Structure, Time, Opinion, Reality, and Yearning. Building on this model, we developed a framework for human-AI collaboration (Figure 1). This framework uses the elements of the data-driven visual storytelling as a baseline and examines the roles humans and AI play in fulfilling these elements. For instance, humans primarily take on roles such as overall direction, emotional depth, and theme design, while AI focuses on rich details, scene expansion, and plot progression. These roles help to illustrate how humans and AI leverage their strengths to contribute to data-driven visual storytelling, offering valuable insights into human-AI collaboration models.

### 3.2 Roles of Humans and AI in data-driven visual storytelling

This study encoded the collaboration modes for each element of the storytelling model to understand the specific collaboration patterns between humans and AI in crafting data-driven visual stories. We first conducted a literature review on methods for storytelling with data [54] and human-AI collaboration in data storytelling [38]. Subsequently, two researchers discussed the findings from the literature to identify which aspects are primarily led by humans and which can be enhanced by AI. Based on an initial framework, the two researchers independently coded the roles of humans and AI across the five storytelling elements. For areas of disagreement, each researcher examined the points of conflict and adjusted the codes, accordingly, ultimately reaching a consensus (as shown in Figure 2). In the structure element, humans primarily oversee the overall direction, including the design of the overarching framework and the control of emotional pacing. At the same time, AI is complementary in enriching and refining details. Humans manage the overall timeline design for time constraints, while AI enhances details, such as displaying stories within a scene along a timeline. In opinion inspiration, humans are responsible for designing the themes, such as presenting stories from multiple perspectives, whereas AI provides contextual support to express these perspectives. Regarding the plot's reality, humans focus on exploring emotional depth, while AI supplements details to enhance realism. Finally, in yearning, humans construct positive values, while AI adds additional emotionally engaging elements to evoke readers' emotional resonance.

### 3.3 Study Protocol of AI News Case

We demonstrate how to create AI ethics narratives through human-AI collaboration by selecting real-world AI ethics events. Lee et al. divided the process of data-driven visual storytelling into three stages: data exploration, story creation, and storytelling [37]. Building on this framework, we structured the narrative construction process for AI news cases as follows: 1) Selecting AI news cases as a data source; 2) Conceptualizing the overall story based on the data; 3) Using generative AI to construct story scenes and enrich story details; 4) Designing an appropriate narrative approach to present the story according to the storyline.

Table 1: Definition of The Five Elements of a Data-driven Visual Storytelling Model

| Element | Definition |
|---|---|
| Structure | A story should have a clear structure, with logical progression and well-defined layers that attract readers and smoothly advance the plot to a satisfying conclusion. |
| Time | The story should have a sense of urgency and a compact timeline, using time frames or deadlines to enhance tension and engagement. |
| Opinion | A story conveys profound ideas or values through its plot and characters, evoking emotional resonance and prompting reflection on life. |
| Reality | The plot should follow logical and emotional truths, even in fictional stories, ensuring believability through reasonable character motivations, event development, and detailed depictions. |
| Yearning | A story should inspire readers to yearn for beauty, hope, or dreams, conveying encouragement, courage, and motivation to take positive action. |

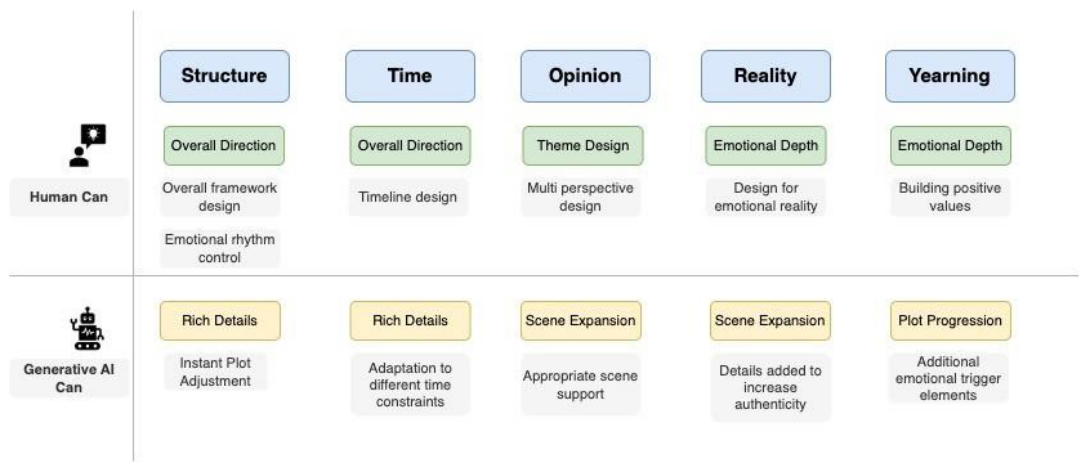

Figure 2: The role division between humans and generative AI in the model of data-driven visual storytelling.

**3.3.1 News Data Selection and Story Ideation.** This study selects a new case from the AI Incident Database. The event summary is as follows: On 14 December 2016, a self-driving Uber vehicle, while stopped at an intersection, suddenly accelerated against a red light and sped toward the opposite junction. The reason for selecting this event is that the field of autonomous driving is a highly scrutinized area closely related to AI ethics and is often associated with numerous fictional AI narratives. Examples of such narratives include autonomous vehicles "choosing" to sacrifice themselves to protect humans at critical moments [47, 58] or autonomous vehicles "thinking independently" and taking control away from humans [7, 25]. We chose this case to demonstrate how to construct AI ethics narratives from the perspective of human-AI collaboration. This case is a representative example highlighting the differences between real-world and fictional AI ethics narratives.

First, we collected all news articles related to this incident, covering the period from 14 December 2016 to 23 March 2017, resulting in 10 different reports. In constructing the story, humans were responsible for determining the overall direction, theme design, and emotional depth. After analyzing all the news articles, we structured the narrative from the perspectives of three key stakeholders.

The researchers primarily completed this step manually through coding. While we acknowledge that generative AI can provide specific suggestions during the story conceptualization process [14], AI ethics experts still construct AI ethics narratives [15]. This remains a critical step in creating compelling AI ethics narratives.

In this study, the researchers' data exploration and story conceptualization process is illustrated in Figure 3. First, all news articles related to the incident were collected and organized chronologically. Based on the characteristics of the news reports and expertise in AI ethics narratives, this paper chose to present the AI ethics incident from the perspectives of different stakeholders to help readers understand the ethical issues involved. Next, statements and actions related to each stakeholder were extracted from the news articles. Finally, the selected texts were refined and structured according to each stakeholder's perspective.

### 3.3.2 Prompt-based Conversational GenAI for Image Generation.

In GenAI, a prompt is an input provided to an AI model to guide its output generation. It provides instructions or context for the GenAI to produce a desired result in the form of text, image, etc. Prompts shape the AI's response by helping the model understand the kind of output expected, such as style and level of detail [57]. In the context of large language models (LLMs), a prompt can be a text or an image, although in some segmentation models, a prompt can be a point, a box, etc. [68]. This study focuses on storytelling and image generation based on AI ethics events in news. Thus, the prompts used in this study are texts. A simple example of a text prompt for image generation can be "Create a landscape of a futuristic city with glowing skyscrapers".

Language nuances in a prompt impact GenAI's output [8]. In order to find the most appropriate prompt, it's usually necessary to fine-tune the prompt [39]. It can happen that this fine-tuning process has to be carried out multiple times. Prompt engineering is the process of designing and crafting the prompt to get desired and accurate responses from a GenAI model. It involves carefully structuring the wording, instructions, and context of a prompt in a way that maximizes the quality, relevance, and precision of the AI's output [8]. Text in an appropriate prompt should be clear, specific, instructive, and able to provide context information [57]. Studies have shown that prompts play a crucial role in vision LLMs [8][68], and prompt engineering is even deemed as "a new literacy in the information age" [41]. An example is shown in Figure 4, in which image (a) is generated using the prompt "Generate an image showing a comparison between Google Waymo self-driving car and Uber: 'Compared to Google/Waymo's overly polite but safe self-

Structured news texts are obtained through the five steps listed in Section 3.3.1. Here is an example of the texts describing the autonomous driving event happening on December 14, 2016, in California [65]: "An Uber (Volvo XC90 SUV) car equipped with self-driving capabilities ran a red light in San Francisco's SOMA neighborhood and almost hit a pedestrian who was running the red light", which is from the perspective of Uber company. Another example is "When the driver was talking to a passenger who took out his laptop, the car suddenly drove forward and passed the red light.", which is from the perspective of a witness.

The structured texts are then input into the GPT-based image creation tool named Image Generator (IG) as prompts. If IG is already able to generate a satisfactory image that illustrates the scene stated in the texts with reasonable details, the image is kept for visualization. But more often, with a first try IG may not be able to generate a satisfactory image. Then, the input prompts have to be fine-tuned (e.g., to add more specific descriptions or details) until a satisfactory image is generated. Another way to improve the image is to use the editing tool provided by IG, which allows for selecting an area to be edited and then adding prompts only for the selected area. Both ways involve prompt engineering. The workflow is shown in Figure 5.

For example, the text "Please generate an image in cartoon style, this is regarding an event of autonomous driving: An Uber (Volvo XC90 SUV) car equipped with self-driving capabilities ran a red light in San Francisco's SOMA neighborhood and almost hit a pedestrian who was running the red light" based on one structured news text listed above is provided as a prompt for IG to create an image. Figure 6 shows the user interface of IG, including the prompt and the generated image. In order to provide a complete prompt, the user just needs to add a simple sentence about the driving cars, Uber cars have also been caught failing to yield to pedestrians'", while (b) created using the prompt "Show the contrast in two parts: above and below. Above shows that the Google Waymo self-driving car behaves politely and safe at a crosswalk. Below shows that an Uber car is failing to yield to pedestrians". The meanings of prompts of (a) and (b) are almost the same. However, the prompt of (b) specifies that the contrast should be made vertically. It disassembles the prompt of (a), which makes the text structure clearer and easier to understand. Therefore, Figure 4(b) presents clear contrast, while (a) looks ambiguous.

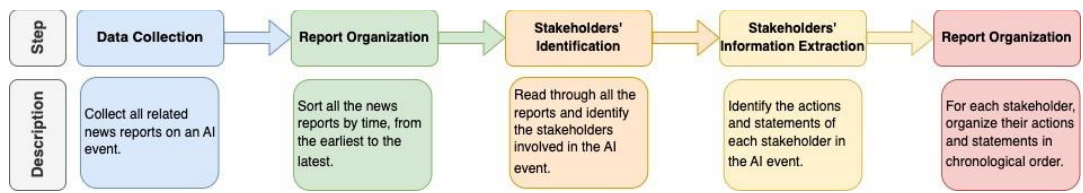

Figure 3: Case study: Data exploration and story ideation for AI news case.

context of the event and the style of the desired image apart from the structured text. IG is able to generate a satisfactory image, depicting the autonomous driving car equipped with self-driving device running into a red light and almost hitting a man. Although details like directions and colors of traffic lights, texts on the car are either mistaken or blurred, the image is still considered satisfactory, as it well depicts the scene in the input prompt with good details.

However, more often IG may not be able to generate a satisfactory image with a first try of a prompt. For example, the prompt is input to IG "generate an image 'San Francisco writer and producer Annie Guas witnessed this incident while riding in a human-driven car. She had just passed a self-driving Uber that swerved into the Van Ness intersection running into a red light and nearly hit her car'". The sentences in the single quotation marks are structured news texts. With a first try, IG generated an image shown in Figure 7(a). Obviously, it did not draw the "writer and producer" and the "human-driven car", and failed in depicting the self-driving Uber "nearly hit her car". Thus, prompt engineering is needed, and in IG this is done in an interactive and conversational way. The prompt "please show explicitly that the Uber nearly hit the human-driven car" is given to IG, and it generates an image shown in Figure 7(b), which is still unsatisfactory, as the "human-driven car" is not drawn. Therefore, the prompt "the Uber is nearly hitting a human-driven car, not a woman" is fed into IG, and an image as displayed in Figure 7(c) is generated accordingly. After a few other tries to fine-tune the prompts, for example, to specify that "It should be one self-driving Uber and one human-driven car", "It's nearly hitting, not already hitting", IG is able to generate a satisfactory image as shown in Figure 7(d).

It is observed that the device on the top of the car in Figure 7(d) looks like a self-driving device. It is stated in the structured news texts and prompts, however, this lady is riding a human-driving car. The editing tool provided by IG allows users to select an area and edit the area with a prompt. With the tool the device on the top of the car is changed, as shown in Figure 8. All other images are generated with this procedure by going through all the structured news texts.

### 3.3.3 Data-driven Visual Storytelling Generation.
Once generative AI enriches the story details and plots following human instructions, we develop the narrative approach in line with the overall direction and thematic design. We adopted the comicboarding method for presentation, aiming to enhance the inclusivity of dialogue among various stakeholders [35, 55]. AI ethics issues, as complex problems involving diverse societal stakeholders, encompass individuals with varying social backgrounds and cognitive levels. Comicboarding, as an inclusive visual analysis tool, enhances emotional resonance and embraces the diversity of audiences from different backgrounds. The comicboarding method has been effectively used in studies to help users understand systems [35], improve public awareness [2, 18, 53], and influence user behavior [31]. As a medium for presenting AI ethics narratives, it facilitates a better public understanding of AI ethics issues, effectively evokes emotional resonance, and prompts deeper reflection [65].

The key characteristic of this new case is that it involves multiple stakeholders, with different event descriptions from each stakeholder's perspective. Using the comicboarding method, we presented the development of the incident in chronological order from the perspectives of four primary stakeholders: eyewitnesses, Uber as a corporation, Uber employees, and the government. These perspectives form the main direction of the AI ethics narrative. When reading a single news report, readers are easily misled by the narrative of that report. For instance, if a reader only encounters a report where Uber's spokesperson claims at a press conference that the driver caused the traffic accident, they may fail to uncover the truth behind the ethical issue. Analyzing all news reports related to the incident and narrating the event from different stakeholder perspectives, we help readers view the event's development holistically, avoiding misdirection by isolated reports. This approach to AI ethics storytelling guides the public in building a correct understanding of AI ethics and facilitates deeper dialogue among stakeholders, ultimately contributing to better policies.

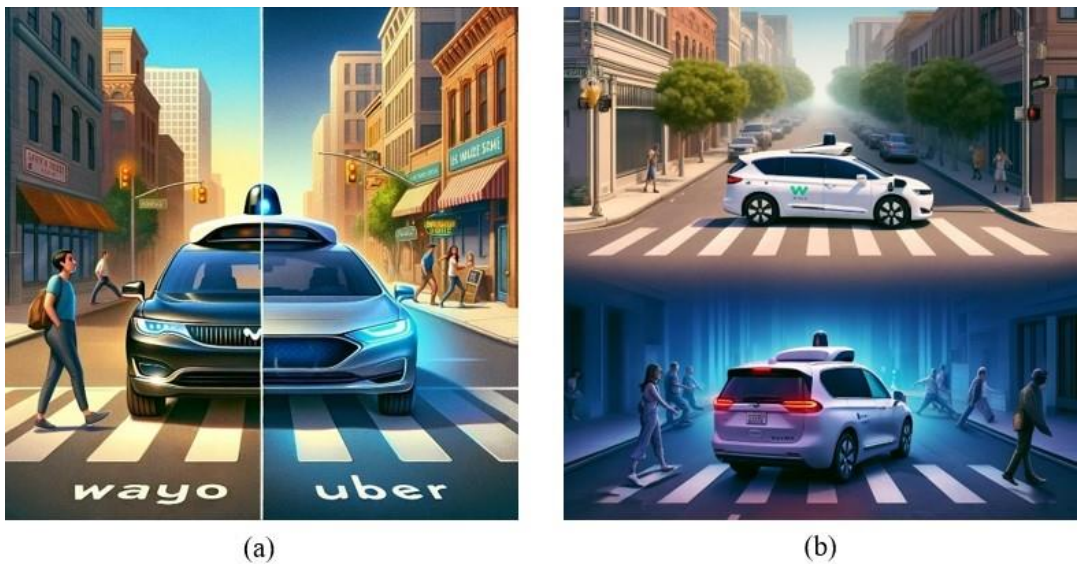

(a)　　　　　　　　　　(b)

Figure 4: An example showing that clear, precise and specific texts in a prompt are the key to generate a satisfactory image.

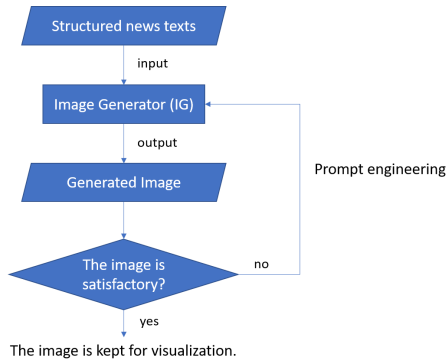

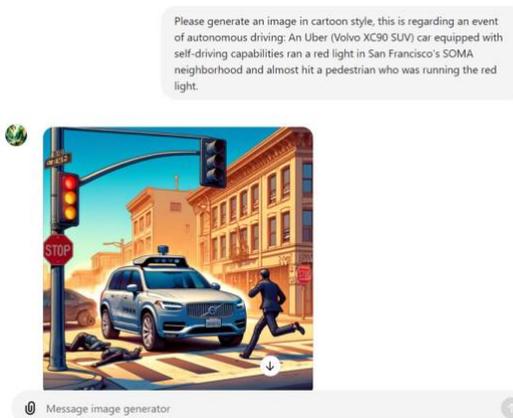

Figure 5: The workflow of generating images using IG with prompts.

Figure 6: The The screenshot of the user interface of IG.

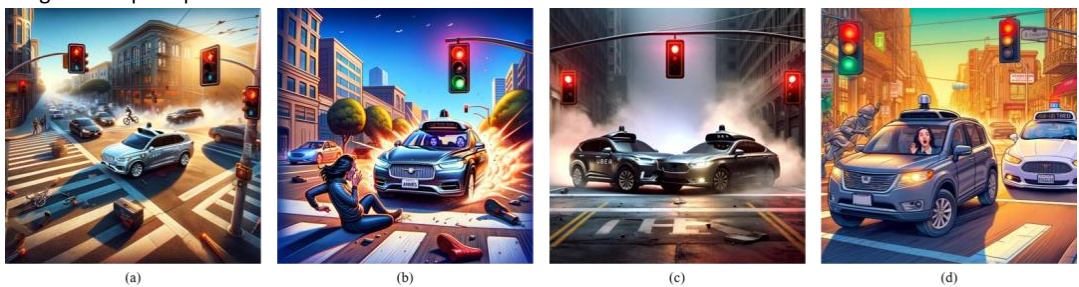

(a)　　　　　(b)　　　　　(c)　　　　　(d)

Figure 7: An example of how multiple attempts at prompt engineering generate a satisfactory image.

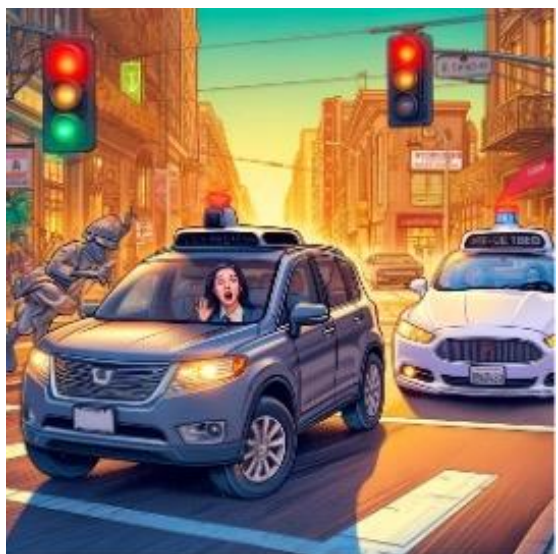

Figure 8: The device on the top of the car is changed with the editing tool in IG.

## 4 Results

In this section, we present a data-driven visual story constructed through human-AI collaboration using a real-world AI news case as the data source (as shown in Figure 9). Our experiment demonstrates two key results: 1) Data-driven visual storytelling facilitated by the human-AI collaboration, and 2) Construction of AI ethics narratives based on real-world cases.

Data-Driven visual storytelling facilitated by the human-AI Collaboration. We proposed a framework for human-AI collaboration in creation of data-driven visual stories using a real-world news case. Based on the analysis of the news content, the story structure adopts an open-ended format. The news event involves multiple stakeholders, each presenting a perspective on the development of the incident. The central question posed in this story is: who should be responsible for an autonomous driving traffic accident? Researchers (humans) organized the actions and statements of different stakeholder and applied generative AI to create visually authentic and impactful narratives. Humans oversee the overall direction, theme design, and emotional depth, while AI enriches the details, expands scenes, and enhances the plot.

In Figure 9, the storyboard depicts an open-ended story structure, unfolding the event sequentially through the perspectives of the four main stakeholders involved in the AI-related news incident. Following the incident, Uber made its first public statement. From Uber's perspective, the company's spokesperson initially claimed that the entire event was caused by human error, blaming the driver, and announcing their suspension as a disciplinary action. Later, Uber argued that pedestrians are not its customers and, therefore, not within its scope of protection. The final frame provides background information on Uber, highlighting the company's greater focus on improving driving efficiency. The second perspective is that of the eyewitnesses. Uber held a second press conference as public opinion intensified, with eyewitnesses emerging to share their accounts. First, the owner of a nearby café at the intersection where the incident occurred claimed to have witnessed the truth. They stated that the vehicle started independently, and the driver was chatting with a friend in the passenger seat. The Uber vehicle suddenly ran a red light at the intersection, narrowly avoiding a collision with an oncoming car. Another eyewitness, Annie, the driver of the oncoming car, stated that as she proceeded through a green light, a vehicle unexpectedly ran the red light and rushed towards her. The third perspective is from Uber's former employees. They revealed that instances of Uber's autonomous vehicles running red lights were not uncommon and that the cars also had issues with failing to yield to pedestrians. The final perspective comes from the government. The government stated that Uber had regulatory issues regarding the application of autonomous driving technology and initiated a legal investigation into the company. This incident garnered significant attention from both the government and Uber. Subsequently, the mayor and the CEO of Uber met to discuss and negotiate safety measures. Creating an effective data story requires integrating data analysis, narrative techniques, and artistic design [24]. In data analysis, researchers determine which data contributes to the story and set analysis goals based on the theme and requirements. At the same time, AI processes the data to quickly generate the necessary visualizations. In artistic design, researchers select visual styles and narrative frameworks that align with the needs of the audience. AI generates richer and more vivid visual results based on aesthetic principles and data structures. In general, humans refine themes and incorporate emotional design to make stories more focused on the audience, while AI generates authentic and engaging visualizations from data. This collaboration creates narratives that are not only emotionally resonant but also visually compelling, allowing readers to experience more prosperous and impactful storytelling.

Construction of AI ethics narratives based on real-world cases. Our study used a new case on autonomous driving to demonstrate the importance of authentic AI ethical narratives in shaping public awareness. In this news case, we first presented a complete AI ethics incident. In the real world, people often only read a single news report about an event. In this case, the first voice in the media came from Uber, which claimed that the human driver caused the accident. If the public only reads this report, they may miss the opportunity to understand the truth. By having experts construct authentic AI ethical narratives, the public can gain a comprehensive understanding of the event [15]. In addition, the open-ended narrative structure presented through the comic panels aligns with current research on issues related to autonomous driving. There are currently no effective policies or regulations to address the ethical challenges posed by autonomous driving. Discussions have often been based on hypothetical 'trolley problem' scenarios, rather than addressing ethical issues in real-world contexts. This paper presented a real-world AI ethics case in the form of comic panels, shifting the narrative of autonomous driving to focus on real-world ethical issues. Exploring the perspectives of Uber, eyewitnesses, company employees, and the government raises open-ended questions to encourage readers' discussion and reflection. As noted by Watson et al. [62], competing narratives contribute to the development of contemporary AI ethics discourse. AI ethics narratives based on real cases can more accurately reflect real ethical issues, avoiding the panic caused by fictional narratives and the neglect of practical yet mundane problems. In addition, constructing AI ethics narratives in the form of comicboardings makes them accessible to readers from diverse backgrounds (including both experts and non-experts). It also enhances emotional resonance and encourages public engagement.

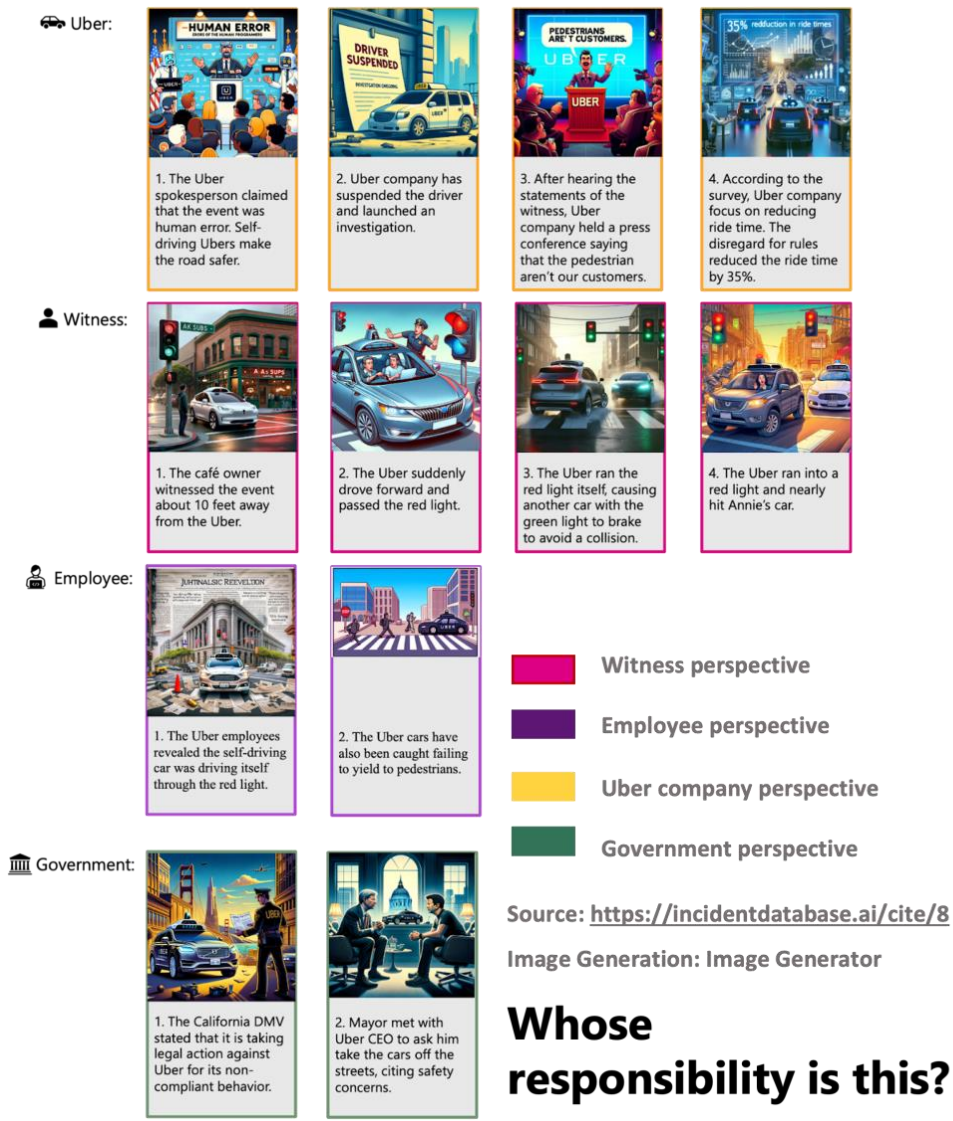

Figure 9: AI ethics narrative based on a real news case as data.

## 5   Discussion

Creating AI ethics narratives based on real-world events is crucial, and data-driven visual storytelling facilitated by the human-AI collaboration offers significant advantages for this purpose. In this study, we proposed a research framework that focuses on how to create compelling data-driven visual stories and explores the role of human-AI collaboration. Building on this framework, we presented the specific roles that humans and AI play in creating such stories. We also considered the implications of this approach, its limitations, and potential directions for future work.

**Exploring human-AI collaboration in creating data-driven visual stories from a new perspective.** Collaboration between humans and machine intelligence is a central topic in human-computer interaction. Generative AI has ushered in a new era for storytellers. As technology continues to advance, collaboration between human creativity and AI-generated content has the potential to redefine the landscape of storytelling [46]. Currently, most research on human-AI collaboration in storytelling focuses on toolcentric explorations. For example, in tool development, Zhang et al. developed StoryDrawer, a co-creation system that supports visual storytelling for children aged 6–10 through collaborative drawing between children and AI [70]. In systematic reviews of tools, Li et al. conducted a comprehensive review of AI-assisted data storytelling tools, presenting a research framework that outlines the stages of storytelling workflows supported by these tools [38].

While tool development and research are undoubtedly critical to human-AI collaboration in crafting data-driven visual stories, the fundamental goal remains how to tell a compelling story. Starting from this goal, this study first analyzes the five key elements of a good storytelling model. Then, it examines how humans and AI can leverage their strengths to achieve this goal collaboratively. Previous research has explored collaboration models where humans create content with AI assistance (human creator + AI assistant) or optimize AI-generated content (AI creator + human optimizer). However, the storytelling process should not be constrained by a single collaboration model. Instead, it is essential to understand and examine complex human-AI collaboration models from multiple perspectives. These models may not be confined to a single stage but could be hybrid. For example, in this study, humans act as creators and leaders when determining the story's structure, while AI serves as an assistant. In achieving realism, AI plays the role of the creator, and humans act as optimizers. Regardless of the collaboration model, the goal is to tell a compelling, data-driven story. As AI technology evolves, exploring how humans and AI can collaborate more effectively will remain an ongoing challenge. Our research aims to provide a new and flexible perspective, driven by the goal of telling compelling stories and leveraging the strengths of humans and AI to achieve effective collaboration.

**The construction of authentic AI ethics narratives is an urgent task.** Narratives, as the original matrix for influencing interpersonal relationships, culture, practices, and learning [50], help people navigate an increasingly complex world [71]. Currently, fictional narratives dominate the field of AI ethics storytelling. These fictional AI ethics narratives cater to the human tendency to be drawn to grandiose and sensational topics, diverting public attention to exaggerated and imaginary risks. More importantly, such narratives often cause people to overlook genuine but seemingly mundane and harmful risks. Chubb et al. argue that current AI ethics narratives are controlled by science fiction and corporate marketing from large technology companies, which distract and mislead the public's understanding and perceptions of AI [15]. They call for creating realistic, nuanced, inclusive stories that embrace diverse voices. However, constructing AI ethics narratives is challenging, many of these calls stop at theoretical discussions. Based on real-world cases, this study demonstrates the process and outcomes of constructing authentic and inclusive AI ethics narratives. Real-world news cases provide a data source for exploring genuine AI ethics issues. Presenting these cases as visual stories using comicboardings can enhance their inclusiveness, making them more accessible to both experts and non-experts.

Furthermore, most real-world AI ethics issues lack definitive or universally correct resolutions. The comicboarding approach, compatible with open-ended story conclusions, encourages public reflection and discussion. It leverages diverse narratives to drive the formulation and improvement of policies [62]. In summary, constructing AI ethics narratives based on real events can redirect public attention to the right areas, enhance public awareness and understanding of AI ethics issues, promote multi-stakeholder discussions and reflections, and ultimately drive the development of policies and regulations.

As a proof-of-concept case study, this paper acknowledges several limitations. First, the roles of human-AI collaboration in crafting data-driven stories could be further detailed at a more granular level. While the current study identifies the roles humans and AI play, it does not specify the exact steps in which humans and AI contribute within the case study. For example, the study lacks detailed illustrations of how humans manage the structure and enhance emotional depth or how AI enriches details and improves realism. Second, due to space limitations, this study validates the practical application of the conceptual framework using only a single real-world case. Such an approach may limit the generalizability of the operational workflow. Our research aims to construct authentic AI ethics narratives. At the current technological stage, the discovery and storytelling of AI ethics issues in real-world cases still require expert guidance. Thus, this process relies heavily on human interpretation and understanding of different cases. The current study primarily demonstrates the potential and advantages of human-AI collaboration in crafting authentic AI ethics narratives. In future work, we aim to classify different AI ethics cases and propose more generalized operational workflows to enable more automation in AI ethics storytelling. For example, by summarizing the specific categories of the five elements of the storytelling model, such as structure (chronological, reverse chronological, flashback), humans can input category instructions when constructing a story, and generative AI can produce stories of specific types, enhancing the automation process of human-machine collaboration.

## 6 Conclusion

Our research demonstrated how to construct AI ethics narratives based on real-world events from the perspective of human-AI collaboration. We proposed a conceptual framework with five elements of data-driven visual storytelling, detailing the roles of humans and AI in each part of the storytelling process. By leveraging the respective strengths of humans and AI, we explore a new human-AI collaboration model for crafting data-driven visual stories. The conceptual framework was implemented using a real-world AI ethics news case, this has helped to address the current gaps in AI ethics storytelling, promote public engagement with technology, and support improving AI policies. In future work, we aim to broaden the analysis of AI news cases and propose a generalized workflow to support the development of authentic, nuanced, and inclusive AI ethics narratives. Additionally, through further user research, we will explore the impact of these AI ethics narratives on enhancing public awareness and understanding of AI ethics.